# Molecular structure of the Discotic Liquid Crystalline Phase of Hexa-peri-Hexabenzocoronene/Oligothiophene Hybrid and their Charge Transport properties

Saientan Bag[1], Vishal Maingi[1], Prabal K Maiti[1*], Joseph Yelk[2], Matthew A. Glaser[2], David M. Walba[3] and Noel A. Clark[2]

[1]Center for Condensed Matter Theory, Department of Physics, Indian Institute of Science, Bangalore 560012, India
[2]Department of Physics, University of Colorado, Boulder-80309, USA
[3]Department of Chemistry and Biochemistry, University of Colorado, Boulder-80309, USA

*Abstract*— **Using atomistic molecular dynamics simulation we study the discotic columnar liquid crystalline (LC) phases formed by a new organic compound having Hexa-peri-Hexabenzocoronene (HBC) core with six pendant oligothiophene units recently synthesized by Nan Hu et al. (N. Hu, R. Shao, Y. Shen, D. Chen, N. A. Clark and D. M. Walba, Adv. Mater. 26, 2066, 2014). This HBC core based LC phase was shown to have electric field responsive behavior and has important application in organic electronics. Our simulation results confirm the hexagonal arrangement of columnar LC phase with a lattice spacing consistent with that obtained from small angle X-ray diffraction data. We have also calculated various positional and orientational correlation functions to characterize the ordering of the molecules in the columnar arrangement. The molecules in a column are arranged with an average twist of 25 degrees having an average inter-molecular separation of ~5 Å. Interestingly, we find an overall tilt angle of 43 degrees between the columnar axis and HBC core. We also simulate the charge transport through this columnar phase and report the numerical value of charge carrier mobility for this liquid crystal phase. The charge carrier mobility is strongly influenced by the twist angle and average spacing of the molecules in the column.**

*maiti@physics.iisc.ernet.in

## INTRODUCTION

Discotic liquid crystals[1,2] (LCs) are typically formed by disc-shaped molecules comprising a central aromatic core with aliphatic chains attached around its edge, organized into phases that are unable to support shear stress applied in one or more directions (i.e., are fluid-like). Discotics exhibit three dimensionally (3D) fluid nematic phases as well as columnar phases in which molecular discs stack on top of each other to form 1D fluid columns which are, in turn, organized on a 2D lattice[3]. The resulting optical, electrical and magnetic properties are anisotropic, i.e. solid−like, due to the global orientational ordering of the disc planes in such structures. The association, stacking, and orientational ordering of their molecular polycyclic aromatic hydrocarbon cores in discotic columnar phase LCs makes them promising candidates for applications in organic electronics[4-8]. Depending on the orientation of the director (the local average orientation of unit vectors normal to the disc planes) with respect to the substrate, DCLCs find applications in various organic electrical devices, wherein charge transport along the columnar stacks is exploited. In this case, if the director is parallel to a planar confining substrate, DCLCs can be used in photovoltaic cells, whereas perpendicular orientation is good for fabrication of field-effect-transistors[4-8]. Discotic columnar LCs (DCLCs) with hexabenzocoronene (HBC) as a aromatic core are the most studied DCLCs, having the highest carrier mobility for charge transport along the columns reported to date[9,10]. The effectiveness of the LC in such devices is dependent on the LC phase structure and fluctuations, and on the morphology of the textural organization of the LC within the device structure. Thus, for DCLCs a key requirement is to get the molecular columns organized as desired relative to the confining surfaces[11].

Exploration of new liquid crystal materials with new features by tuning the periphery, shape of the core and the nature of attached group have been going on through past several years. A recent report[12] by Nan Hu et al. of new HBC based columnar liquid crystal with six pendant quadra-3-hexylthiophene units attached to the core with long alkyl chain draws our attention. Henceforth, we will denote this molecule as W732. The material is reported to be easily functionalized[13] and greatly applicable in organic electronics[14-16]. The new material can self assemble[17] in columnar liquid crystal state and shows thermodynamically stable enantriotropic columnar phase over a broad temperature range in spite of having bulky heterocyclic aromatic unit at the end of the long tail[12]. The most striking feature of the newly reported LC phase is its ability to provide well aligned cells controllable by electric field[12].

There are many methods[18-22] for aligning the substrate in homeotropic or parallel alignment, but none of the method gives a high quality alignment of the sample. Furthermore, except very few cases the liquid crystal phases are rarely reported to have electric field responsive behaviour[23-26] and electric field controlled alignment.

Realizing the potential of the HBC based liquid crystalline phase it is necessary to have more microscopic insight about the columnar phase which is not easily obtained from the experiment. Molecular dynamics (MD) simulation[27-29] being a powerful tool to extract microscopic information about the phase is a natural choice in this situation. MD simulation has been extensively used[30-33] in recent years to study the various equilibrium properties of the DCLCs . Recently Andrienko et. al.[29] have reported MD simulation study of the discotic phase formed by alkyl-substituted HBC mesogens where both the static and dynamic properties of the DCLC was explored. Beltrán[34] et. al successfully employed MD simulation on star shaped discotic liquid crystals which give rise to two different nanosegregated architectures. In a very recent report, Busselez et. al.[35] studied the discotic columnar liquid crystals of model Gay-Berne system both in the bulk as well as nanoconfined states using molecular dynamics simulation.

The properties of this newly reported DCLC are very much attractive for their potential application in molecular electronics as has been discussed earlier. Use of this compound for molecular electronics application requires a quantitative understanding of charge carrier mobility through the column of this molecule in the discotic phase, which is lacking at the moment. Charge carrier mobility calculation in wide variety of organic semiconducting material has been reported in the past. A theoretical calculation of hole mobility in Oligoacene was studied by Deng et al.[36] in 2004. Köse et. al.[37] have reported charge carrier mobility simulation in dendrimers . Charge carrier mobility in HBC based DCLCs was first studied by Kirkpatrick et. al.[38] using semi classical Marcus–Hush formalism. Haverkate et. al.[39] have reported the effects of conformation, defects and dynamics on charge transport properties of DCLCs. In a recent report, charge transport property of the liquid crystal heterojunction has been analyzed by Idé et.al.[40]. Using semi classical Marcus–Hush formalism in this paper, we also give a quantitative estimate of the charge carrier mobility in the columnar phase of W732.

The rest of the paper is organized as follows: In section I we give the details of the model building and the MD simulation of the discotic phase of W732. Section II is devoted to the discussion of charge transport simulation. In section III we summarize our work with concluding remarks.

## I. MOLECULAR DYNAMICS SIMULATION

### I.A. MODEL AND COMPUTATIONAL DETAILS

The molecular structure of the model simulated is shown in figure 1(a). The structure consists of flat aromatic HBC core with six alkyl chain at the end of which pendant quadra-3-hexylthiophene units are attached. Initially the molecule was fragmented in parts and each part was geometry optimized in Gaussian[41] with HF/6-31G(d) as the basis set. During optimization, charges were calculated using electrostatic potential (ESP) method. Then the optimized fragments were added together using Dendrimer builder ToolKit[42] to get the 3-d model of the molecule for further studies. The Antechamber[43] module of the AMBER[44] package with the GAFF[45] force field and ESP charge (calculation of charge fitted to electrostatic potential) was further used to prepare the system for simulation.

A system of 160 molecules, arranged in 16 columns of 10 molecules each was simulated. We have done the simulation with two different initial conditions. In one case, the columns were arranged on a hexagonal lattice with lattice spacing equal to the experimental value of 52.7 Å[12]. In another case we started with an initial lattice constant of 43 Å. Molecular organization in the initial hexagonal closed pack structure with lattice constant of 52.7 Å and 43 Å are shown in fig. 1(b) and 1(c) respectively. The terminal aromatic parts from molecules belonging to adjacent columns are inter-digitated giving rise to a closed pack structure in both the cases. In both the cases, the simulated system contained 143,040 atoms.

After an initial energy minimization, the system was heated slowly from 10 K to 357 K at a constant pressure of 1 bar. A production run of 100-200 ns was then performed at 357 K and at a pressure of 1 bar. At 357 K the system is in the columnar (COL) liquid crystal phase as seen in experiment and well characterized by the synchrotron small angle scattering X-ray experiments[12]. Temperature and pressure were kept fixed using a Berendsen weak temperature coupling and pressure coupling method[46]. During the NPT run simulation box angles were fixed at $90^0$ and length of the simulation box was allowed to vary independently in each direction. The bonds involving hydrogen were kept constrained using the SHAKE algorithm. This allowed us to use an integration time step of 2 fs. All calculations were performed with parallel version of PMEMD[44].

**I.B. RESULTS AND DISCUSSION**

To monitor the equilibration during the MD simulation, we plot the lattice constant of the system as a function of the simulation time in figure 2. We find that the system equilibrates at a lattice constant of ~46.5 Å within 50 ns. We have repeated the simulation starting two different initial configurations and find that in both the cases the lattice constant converges to ~46.5 Å. It is worth mentioning here that due to large molecular weight of W732, it is very difficult to get adequate conformational sampling of the molecules within the column at such high density even after 100-200 ns long MD simulation. One way to circumvent this problem is to start with the low density configuration and let the system evolve to generate dense columnar configuration as seen in the experiment. However, starting from low density configuration to achieve dense columnar configuration as in experiment will be computationally challenging at this moment.

A representative equilibrated simulation snapshot of the system with 160 molecules in 16 columns at *T=357K* is shown in figure 3(a) .To see the global arrangement of molecule clearly only the HBC cores are shown in the figure 3(b) (top view) and figure 3(c) (side view) . The structure is a clearly hexagonal columnar with HBC cores tilted with respect to the columnar axis. We now quantify the nanoscale ordering of the phase in various ways and offer in the following sections molecular level details of the structures formed.

### I.B.1. Positional order

To quantify the in-plane order in each column we represent each column of the system by their centre of mass and plot on a plane as shown in figure 4. It again shows nice hexagonal structure (figure 4) with a lattice constant of ~46.5 Å. The value of the simulated lattice constant is smaller than those provided by X-ray scattering [12], suggesting that the molecules in our simulation are somewhat more tightly packed[47] and ordered. It is worth mentioning that Andrienko et. al.[47] also reported smaller lattice constant in their simulation compared to the experimental lattice parameter. To have more quantitative comparison with the experimental data we have also calculated the X-ray diffraction (XRD) pattern to be expected from this 2-d cross section of the hexagonal lattice. The simulated pattern is shown in figure 5(a) along with the experimental XRD pattern shown in Fig. 5(b). We find very similar XRD peaks in the simulation as in experiment with the prominent peak at q =0.157 Å$^{-1}$ corresponding to a d-spacing of 40 Å, q =0.273 Å$^{-1}$ corresponding to a d-spacing of 23.01 Å. The difference in the peak positions and relative intensities between the simulated and experimental XRD pattern is due to the fact that the simulated lattice constant of the phase is slightly smaller than the experimental value. Our simulation results confirm arrangement of column on regular hexagonal lattice.

Characterization of the columnar order of the mesophase is necessary for the development of various applications to molecular electronics, as we have already mentioned. We characterize columnar order by measuring positional and orientational order of the molecules in a column. To extract information about the positional order of the molecules parallel to the columnar axis we calculate two types of pair correlations function[48] defined below.

$$g^{cc}(r_\parallel, r_\perp) = \frac{1}{4\pi} \int_0^{2\pi} d\varphi \, g^{cc}(r_\parallel, r_\perp, \varphi). \qquad (1)$$

$$g(r_\parallel) = \int dr_\perp \, g^{cc}(r_\parallel, r_\perp). \qquad (2)$$

Here $r$ is the vector connecting the centre of mass of any pair of molecules along the column, and $r_\parallel$ and $r_\perp$ (shown in the upper left corner of inset of figure 6(b)) denote the component of $r$ parallel and perpendicular to the columnar axis i.e. $r_\parallel = r.\hat{z}$ and $r_\perp = |r - (r.\hat{z})\hat{z}|$ ($\hat{z}$ is the unit vector along columnar axis). $\varphi$ is the azimuthal orientation of $r$ around the columnar axis (along z axis of the simulation box).

The correlation function $g(r_\parallel)$ is shown in figure 6(a). The observation that the peaks are equally spaced at 5 Å interval suggests that the average vertical separation between two neighboring HBC cores plane is ~5 Å. The decay of peak height is the signature of decay in correlation (positional) between molecules with increasing distance along the columnar direction. Figure 6(b) shows a contour plot of the calculated pair correlation function $g^{cc}(r_\parallel, r_\perp)$. The contour plot can be better understood with the help of figure 6(a). We find that the height of the peaks in $g^{cc}(r_\parallel, r_\perp)$ decrease in amplitude with increasing magnitude of $r$ and on an average the $r_\perp$ value is increasing as we increase the $|r_\parallel|$ values. While the decrease of peak amplitude is due to the decay in correlation (positional) as mentioned above, the increase of $r_\perp$ value indicates the kind of arrangement shown schematically in figure 6(c). To verify whether the bending of the columns as shown in Fig. 6(c) is due the constraint that the periodic unit cell is kept to be orthorhombic during the NPT MD simulation, we have performed another set of simulation in a triclinic unit cell using Rahman-Parinello barostat. No significant changes in the columnar organization were found.

### I.B.2. Orientational order

The parameter that governs the charge transport along the column in a significant way is the transfer integral[47,49]. The transfer integral being intimately related to molecular overlap is strongly dependent on the relative orientation of the neighbouring molecule in a column. To get qualitative understanding about the transfer integral and the way they influence charge transport, we calculate the probability distribution of the difference of the azimuthal rotation angle (twist) between the nearest neighbour molecules belonging to the same column and the probability distribution of the angle (relative tilt) that the normal of the HBC cores of two neighbouring molecules (belonging to same column) make with each other. From figure 7(a) an average twist of 25 degrees is noted. Figure 7(b) indicates that the HBC core of the neighbouring molecule are not parallel but these make an angle of 5 degrees with respect to each other.

To complete the description of molecular orientational order, we also measure the tilt angle that the normal of the HBC core of a molecule makes with the column axis. The probability distribution of the tilt angle is shown in figure 8. It is evident from figure 8 that the HBC cores mostly maintain a tilt angle of 43 degrees with respect to the columnar axis. We have also measured the average tilt of the HBC cores of the individual columns to check for any correlation in tilt direction within the columns. No correlation in tilt direction was found.

**I.B.3.Nanophase segregation**

In the experiment, a diffuse scattering peak at 4.47 Å was assigned to a nanophase segregated phase of the pentacyclic aromatic oligothiophene-triazene (thiophene) residues. In our simulation however, we find no clear evidence of the nano-phase segregation of the thiophene unit from the density distribution of the thiophene residues.

**I.B.4. Dielectric Anisotropy**

Electro-optic response of the liquid crystal being the most striking features of it demands proper understanding. Since the phase is not ferroelectric in nature it was proposed that dielectric anisotropy[50] may be responsible for the observed switching from parallel to homeotropic alignment under application of electric field. Dipolar reorientation of the triazole-oligiothiophene (dipolar in nature) chain was thought to be responsible for the dielectric anisotropy in the molecule. To verify this we check how the oligothiophene-traizene groups are oriented with respect to the columnar axis. We calculate the cosine of the angle made by the long axis of the aromatic tails (shown in the inset of figure 9) with respect to the columnar axis, with the resulting histogram shown in figure 9. It is evident from the figure that the thiophene units are mainly oriented perpendicular to the columnar axis. This validates the proposed reason[12] behind positive dielectric anisotropy.

**II. CHARGE CARRIER MOBILITY**

To study the charge carrier mobility along the column of the discotic mesophase we follow the semi-classical Marcus-Hush formalism[51] which has been successfully used[47,52,53] previously to predict the charge carrier mobility in various similar kind of discotic phases. According to the formalism the charge transfer rate $\omega_{ij}$ from $i^{\text{th}}$ molecule to the $j^{\text{th}}$ molecule is given by

$$\omega_{ij} = \frac{|J_{ij}|^2}{\hbar} \sqrt{\frac{\pi}{\lambda kT}} \exp\left[-\frac{(\Delta G_{ij} - \lambda)^2}{4\lambda kT}\right]. \qquad (3)$$

$J_{ij}$ is the transfer integral, defined as

$$J_{ij} = \langle \phi^i | \hat{H} | \phi^j \rangle. \qquad (4)$$

Here $\phi^i$ and $\phi^j$ is the diabatic wave function localized on $i^{\text{th}}$ and $j^{\text{th}}$ molecule respectively. $\hat{H}$ is the Hamiltonian of the two molecules system between which the charge transfer takes place. $\Delta G_{ij}$ is the free energy difference between two molecule. $\lambda$ is the reorganization energy , $\hbar$ is the Plank's constant, k is the Boltzmann's constant, and T is the temperature.

The work flow[52] of our charge transport simulation is as follows: i) We use fully atomistic MD simulation technique to get the equilibrium morphology of the system ii) With this morphology we calculate the transfer integral for each neighbouring pairs in a column and the all other quantities appearing in the rate expression iii) Once the rate $\omega_{ij}$ is known for all neighbouring pairs kinetic Monte Carlo[52,54] method is used to simulate charge carrier dynamics and calculate mobility.

The details of molecular dynamics simulation has already been discussed in section I. Once equilibrium morphology is known from MD simulation the side chains of the HBC core are ignored for further studies since the side chains have no role in charge transfer along the column. Note that we have only considered the charged transport through the HBC core and ignored the possibility of transport through the thiophene units.The HBC cores of the molecules self-assembles to form a well ordered columnar structure which gives rise to a narrow distribution in the transfer integral distribution (Figure 10). As a result, mobility along the HBC core of the column is very high. On the other hand no self-assembly was found for the terminal oligothiophene groups. They are rather in disordered state. This kind of disorder kills the mobility drastically. Now each aromatic core is replaced by rigid energy minimised structure[52,55] with same axial and torsional orientation as obtained from MD simulation to get rid of the bond length fluctuation introduced by thermal fluctuation during MD simulation[55]. The highest occupied molecular orbital (HOMO) and lowest unoccupied molecular orbital (LUMO) are used as diabatic wave function to calculate $J$ for hole and electron transfer respectively. In the case of HBC, due to molecular symmetry both HOMO and LUMO are doubly occupied , so we calculate four terms[56,57] $\langle HOMO^i | \hat{H} | HOMO^j \rangle$, $\langle (HOMO-1)^i | \hat{H} | HOMO^j \rangle$, $\langle HOMO^i | \hat{H} | (HOMO-1)^j \rangle$, and $\langle (HOMO-1)^i | \hat{H} | (HOMO-1)^j \rangle$, and the $J$ (for hole transport) is the root mean square of these four terms[56,57]. To calculate the four terms we use the density functional theory (DFT) method with Becke three-parameter Lee-Yang-Parr (B3LYP) hybrid functional, and 6-311G basis set using Gaussian[41] programme and the code available in VOTCA-CTP[55] module .The reorganization energy $\lambda$ can be decomposed in two parts, inner sphere reorganization energy and outer sphere reorganization . Inner sphere reorganization energy takes care of the change in nuclear degree of freedom as charge transfer takes place from molecule $i$ to molecule $j$ and can be defined as

$$\lambda_{ij}^{int} = U_i^{nC} - U_i^{nN} + U_i^{cN} - U_i^{cC} \qquad (5)$$

Where $U_i^{nC}(U_i^{cN})$ is the internal energy of neutral (charged ) molecule in charged (neutral) state geometry and $U_i^{nN}(U_i^{cC})$ is the internal energy of neutral (charged) molecule in neutral (charged) state geometry. Outer sphere reorganization energy is that part of the reorganization energy that takes into account the reorganization of the environment as the charge transfer takes place. If the charge transfer is not taking place in polarisable environment outer sphere reorganisation energy can be neglected[52,55] and the inner sphere reorganization energy is the only contribution to the total reorganization energy. The value of $\lambda$ used in our simulation is 0.1 eV (for hole transport) which was calculated using same level of theory (b3lyp/6-311g) as used in the calculation of transfer integral. . To calculate the free energy difference $\Delta G_{ij}$, only the contribution from the external electric field $F$ was taken into account and $\Delta G_{ij}$ takes the form $F.d_{ij}$ under this assumption. Here $d_{ij}$ is the displacement vector between the centre of mass of $i^{th}$ and $j^{th}$ molecule. The polar contribution, the contribution due to internal energy difference and the electrostatic contribution to the free energy difference are neglected without any significant error because the environment is not polarisable , the two molecules are identical and the molecules (being parallel to each other) are facing almost same electrostatic environment respectively[52,55]. Now $\omega_{ij}$ is calculated for all neighbouring pairs and charge transport dynamics is simulated using kinetic Monte Carlo (MC) method . For the kinetic MC, an in house code was developed with following algorithm[52,55]. We take a column (taken from the snapshots of the MD simulation) and assign an unit positive charge to a molecule $i$ belonging to that column. At this point we initialize the time as $t = 0$. Now the waiting time $\tau$ of the charge is calculated according to the relation

$$\tau = -\omega_i^{-1} \ln(r_1) \qquad \omega_i = \sum_j \omega_{ij} \qquad (6)$$

and time is updated as $t = t + \tau$ . Where $j$ is the index of the nearest neighbours (in our case two nearest neighbours) of the molecule $i$ and $r_1$ is a random number between 0 to 1. The charge can hop to two positions (two nearest neighbour sites). To decide where the charge will hop we call a random number $r_2$ between 0 and 1 and check the condition $\frac{\omega_{ik}}{\omega_i} < r_2$ ($k$ is the index of any one of the two nearest neighbours of $i^{th}$ molecule). If the condition is satisfied then the charge will be moved to $k^{th}$ position, otherwise it will move to the other neighbour. Now the position of the charge is updated and the above process is repeated.

The simulation was repeated with different column (as in the snapshots obtained from MD) with different initial positioning of the charge on that column. The charge carrier mobility was determined[55] from the average charge velocity given by

$$\langle v \rangle = \mu F \qquad (7)$$

$\langle v \rangle$ is the average charge velocity of the charge, $\mu$ is the mobility and $F$ is the applied electric field. In our case average velocity was calculated from unwrapped charge displacement (since periodic boundary condition was used) divided by the total simulation time $t$.

The numerical value of charge carrier mobility was found to be 0.23 cm$^2$V$^{-1}$s$^{-1}$ at T=300K at electric field strength of $10^7$ V/m. This value of the charge carrier mobility can be compared to the mobility of 0.13 cm$^2$V$^{-1}$s$^{-1}$ for PhC$_{12}$ systems[52]. To relate the calculated charge carrier mobility to the molecular ordering present in our system we present a plot of probability distribution function of logarithm of square of transfer integral in figure 10. The transfer integral being related to molecular overlap strongly depends on the relative position and orientation of neighbours. Two points to note from this graph: position of the peak and the width of the peak. The occurrence of the peak at a very low value of $J$ can be very easily associated with the molecular ordering present in our system. As described in section I.B.2. the nearest neighbour molecules are mostly twisted at an angle of $25^0$ with respect to each other and it is well known in the literature[47] that 25 degree twist gives minimum value of $J$ for HBC core while maximum value of $J$ occurs at a twist angle of 60 degree. Moreover average distance between two nearest neighbour molecules in our system is 5 Å which gives rise to lower value of $J$ as it varies as $\exp(-2.2z/\text{Å})$ with the distance $z$ between two molecule [47]. With this low value of $J$ we still get appreciable value of charge carrier mobility because of the lower width of the peak . For 1d charge transport, charge carrier mobility is limited by the low value of transfer integral available in charge transport pathway that's why width of the distribution is important: sharper the peak, larger the value of mobility. Lower width of the distribution function indicates better arrangement (both positional and orientational) of molecule along the column.

## III. SUMMARY AND CONCLUSION

In summary, we have simulated the discotic columnar liquid crystalline phase of Hexa-peri-Hexabenzocoronene-oligothiophene Hybrid. The simulated phase with 10 column and 160 molecules showed nice arrangement of the column on a hexagonal lattice with lattice constant 46.5 Å. We have calculated various positional and orientational correlation functions between the molecules both along and perpendicular to the director. The average vertical separation between two HBC plane was found out to be 5 Å . We have also found that the HBC core of the neighbouring molecules in a column was mostly twisted at an angle of 25 degree with respect to each other while the core of each molecule maintain an average angle of 43 degree with respect to the director. No indication of nano-phase segregation was found in the simulated phase. We have shown evidences for the molecular dielectric anisotropy which is responsible for the electric-field responsiveness of the phase. We have also calculated the value of charge carrier (hole) mobility using KMC and find the mobility to be 0.23 cm$^2$V$^{-1}$S$^{-1}$ in this phase. We hope that electric field responsiveness[12] as well as an appreciable value of mobility will make the material a promising candidate for molecular electronics.

## ACKNOWLEDGEMENTS

We acknowledge financial support from the Department of Science and Technology (DST), India.

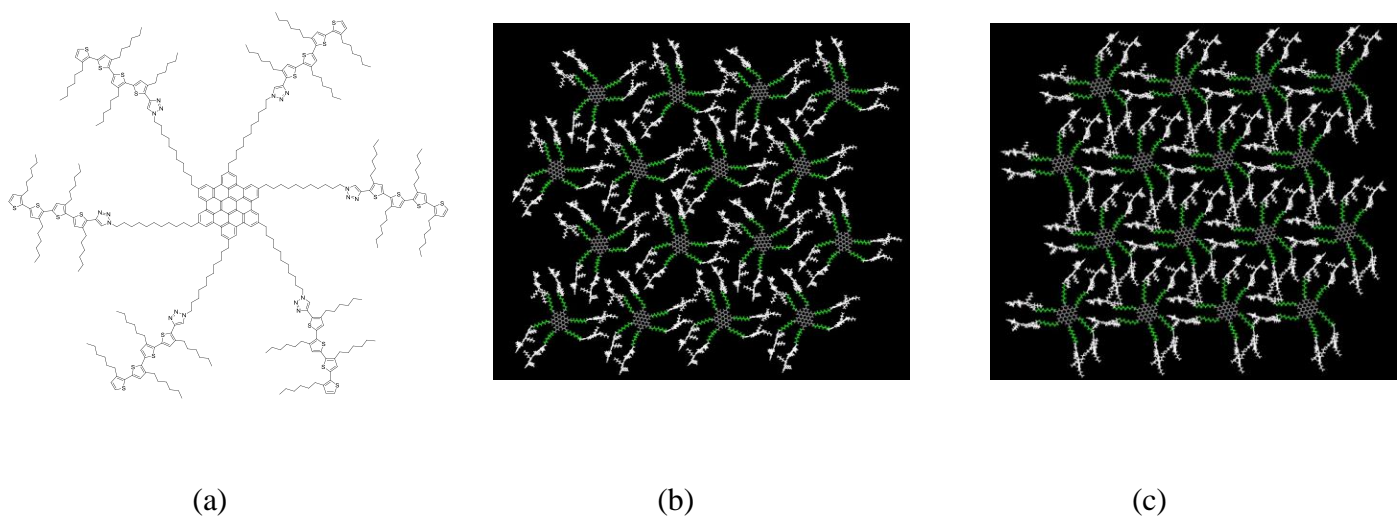

(a)  (b)  (c)

Figure 1: Molecular structure of the model used for simulation (a). The structure consists of flat aromatic HBC core with six alkyl chain at the end of which pendant quadra-3-hexylthiophene units are attached (a). Molecular organizations in the initial hexagonal closed pack structure with a lattice constant of 52.7 Å (b) and 43 Å (c).

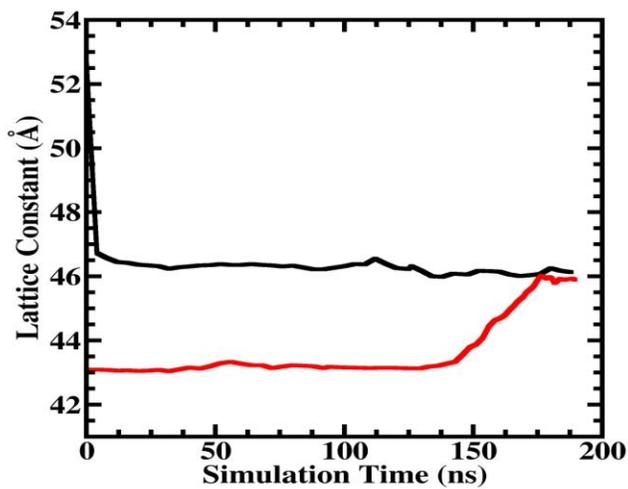

Figure 2: Lattice constant of the systems as a function of simulation time. Two different system was prepared initially with lattice constant 52.7 Å (black line) and 43 Å (red line) respectively. Both the systems at equilibrium converge to a lattice constant value of ~46.5 Å.

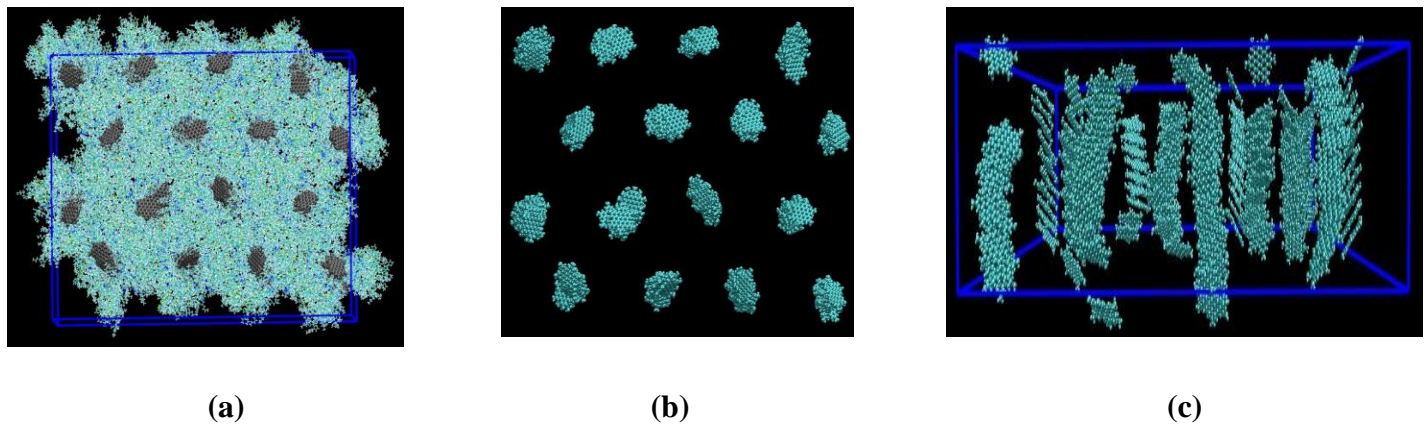

**(a)**                    **(b)**                    **(c)**

Figure 3: An equilibrated representative simulation snapshot of the system with 16 columns and 160 molecules (a). Same snapshot with only HBC cores shown in (b) (top view) and (c) (side view). Note the hexagonal lattice structure in (b) with HBC cores tilted with respect to the columnar axis (c).

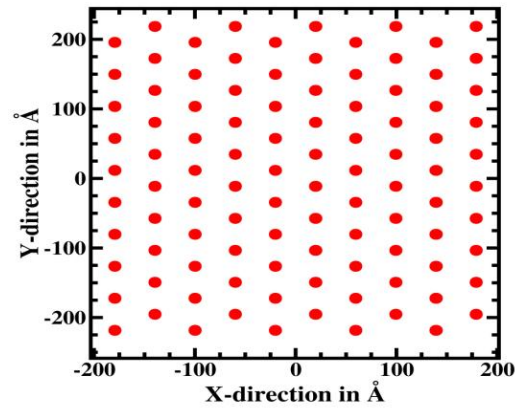

Figure 4: Schematic representation of the simulated system (periodically extended) on a plane. Centre of mass of a column is represented by red dots in figure. This 2-d cross section was used to simulate XRD pattern to be obtained from the LC phase.

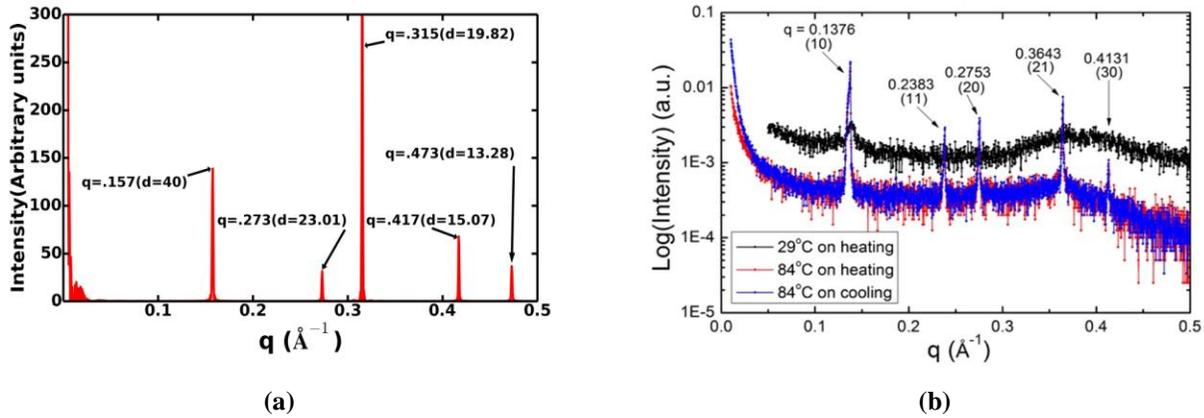

**(a)**                                              **(b)**

Figure 5: Simulated XRD pattern (a) of the 2d crystal (shown in figure 4.) Experimental XRD pattern (b) . Note very similar XRD peaks as in experiment with the prominent peak at q =0.157 corresponding to a d-spacing of 40 Å, q = 0.273 corresponding to a d-spacing of 23.01 Å.

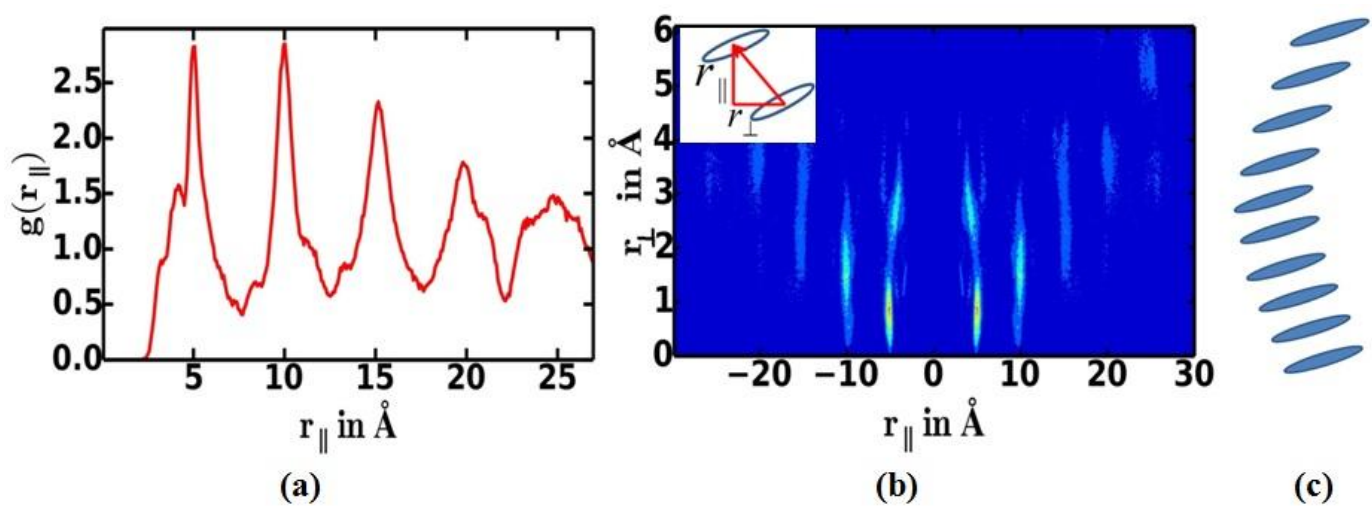

**(a)**                                    **(b)**                          **(c)**

Figure 6: The pair correlation function $g(r_\parallel)$ (a). Note the peaks are equally spaced at 5 Å interval. The peak height decays with increasing distance (a). Two dimension pair correlation function $g^{cc}(r_\parallel, r_\perp)$ (b). The peaks in $g^{cc}(r_\parallel, r_\perp)$ decrease in amplitude with increasing magnitude of $r$ and on an average the $r_\perp$ value is increasing as we increase the $|r_\parallel|$ values (b).Definition of $r_\parallel$ and $r_\perp$ is illustrated in the inset (upper left corner). (b). Schematic representation of the molecules in a column (c).

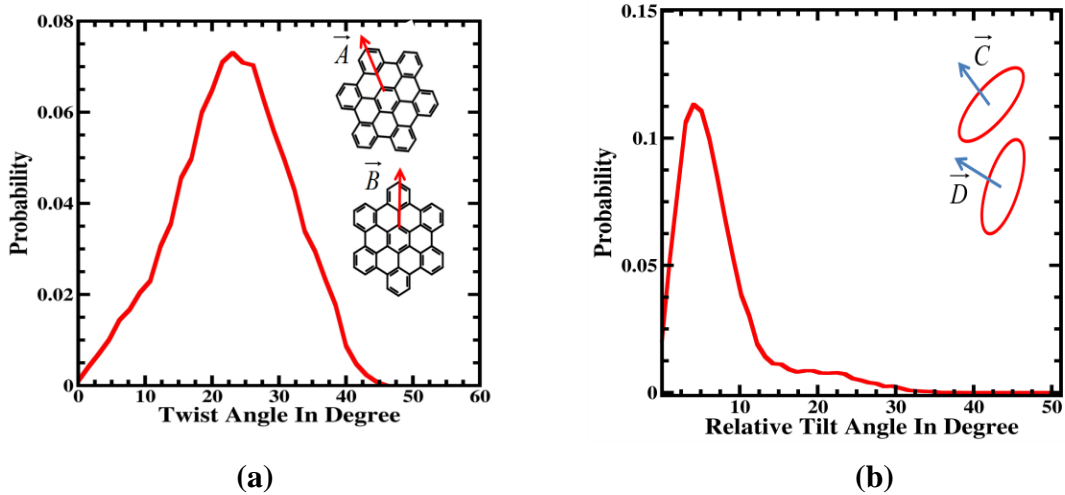

**(a)**  **(b)**

Figure 7: Probability distribution function of the twist angle between neighboring molecules in a stack(a) . The angle between the vectors **A** and **B** (see inset) is defined as twist angle (a). The curve shows peak at an angle of ~25 degrees (a). Probability distribution of the relative tilt angle between neighboring molecules in a stack (b). **C** and **D** (see inset) are the normals of the HBC core. The angle between the vectors **C** and **D** is defined as relative tilt angle (b).

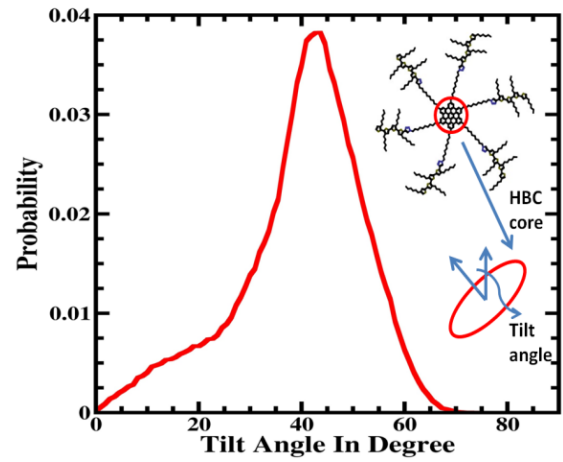

Figure 8: Probability distribution function of the angle of tilt of HBC core relative to average columnar axis. The inset illustrates the definition of tilt angle. The tilt distribution is peaked around 43 degrees.

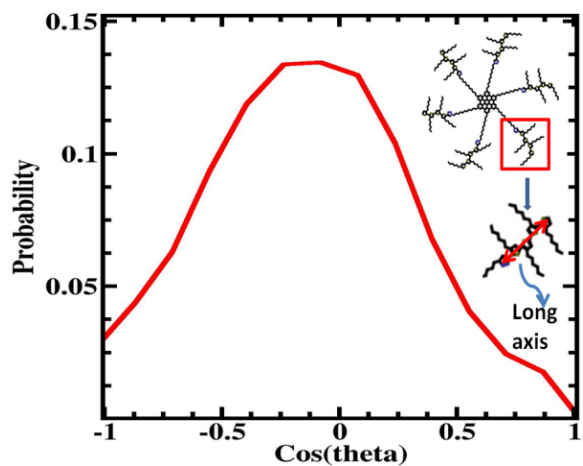

Figure 9: Probability distribution of the cosine of the angle (theta) between the long axis of aromatic oligothiphene-triazene with columnar axis. Inset illustrates the definition of long axis. The oligothiophene triazene are mostly normal to the columnar axis.

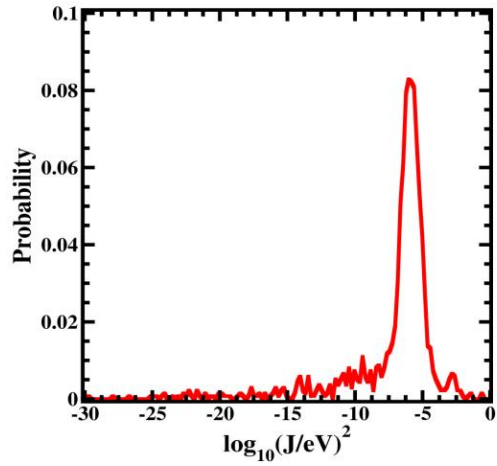

Figure 10:  Probability distribution of the transfer integral. The distribution function features a sharp peak at a value of -6.